\documentclass[a4paper]{spie}  

\usepackage[a4paper, left=1.925cm, right=1.925cm, bottom=4.94cm, top=2.54cm]{geometry}
\usepackage{float}
\usepackage[squaren,textstyle]{SIunits}
\addunit{\arcsec}{as}
\addunit{\parsec}{pc}
\usepackage{amsmath,amsfonts,amssymb}
\usepackage{graphicx}
\usepackage[colorlinks=true, allcolors=blue]{hyperref}
\usepackage{epstopdf}

\title{The metrology system of the VLTI instrument GRAVITY}

\author[a]{Magdalena Lippa}
\author[a]{Stefan Gillessen}
\author[b]{Nicolas Blind}
\author[c]{Yipting Kok}
\author[a]{\c{S}enol Yaz{\i}c{\i}}
\author[a]{Johannes Weber}
\author[a]{Oliver Pfuhl}
\author[a]{Marcus Haug}
\author[a]{Stefan Kellner}
\author[a]{Ekkehard Wieprecht}
\author[a]{Frank Eisenhauer}
\author[a]{Reinhard Genzel}
\author[a]{Oliver Hans}
\author[a]{Frank Hau{\ss}mann}
\author[a]{David Huber}
\author[a]{Tobias Kratschmann}
\author[a]{Thomas Ott}
\author[a]{Markus Plattner}
\author[a]{Christian Rau}
\author[a]{Eckhard Sturm}
\author[a]{Idel Waisberg}
\author[a]{Erich Wiezorrek}
\author[d]{Guy Perrin}
\author[e]{Karine Perraut}
\author[f]{Wolfgang Brandner}
\author[g]{Christian Straubmeier}
\author[h]{Antonio Amorim}
\affil[a]{Max Planck Institute for Extraterrestrial Physics (MPE), Giessenbachstr. 1, 85748 Garching, Germany}
\affil[b]{Universit\'e de Gen\`eve, 24 rue du G\'en\'eral-Dufour, 1211 Gen\`eve 4, Switzerland}
\affil[c]{The University of Western Australia, 35 Stirling Highway, Perth WA 6009, Australia}
\affil[d]{Observatoire de Paris/LESIA, 61 Av. de l'Observatoire, 75 014 Paris, France}
\affil[e]{IPAG, 414 Rue de la Piscine, Domaine universitaire, 38 400 Saint Martin d'H\`eres, France}
\affil[f]{MPIA Heidelberg, K\"onigstuhl 17, 69117 Heidelberg, Germany}
\affil[g]{Univ. Cologne, Z\"ulpicher Str. 77, 50937 K\"oln, Germany}
\affil[h]{SIM FCUL, Edif\'icio C8, gab 8.5.12, 1749-016 Lisboa, Portugal}

\authorinfo{Further author information: \\Magdalena Lippa: E-mail: lena@mpe.mpg.de}

\begin{document} 
\maketitle

\small Copyright 2016 Society of Photo-Optical Instrumentation Engineers. One print or electronic copy may be made for personal use only. Systematic reproduction and distribution, duplication of any material in this paper for a fee or for commercial purposes, or modification of the content of the paper are prohibited.
\normalsize

\begin{abstract}
The VLTI instrument GRAVITY combines the beams from four telescopes and provides phase-referenced imaging as well as precision-astrometry of order $\unit{10}{\micro\arcsec}$ by observing two celestial objects in dual-field mode. Their angular separation can be determined from their differential OPD (dOPD) when the internal dOPDs in the interferometer are known. Here, we present the general overview of the novel metrology system which performs these measurements.  The metrology consists of a three-beam laser system and a homodyne detection scheme for three-beam interference using phase-shifting interferometry in combination with lock-in amplifiers. Via this approach the metrology system measures dOPDs on a nanometer-level.
\end{abstract}


\section{INTRODUCTION}
\label{sec:intro}  

GRAVITY combines the beams from four telescopes interferometrically and provides high-resolution imaging as well as narrow-angle astrometry together with spectroscopic and polarimetric capabilities. For phase-referenced imaging and precision-astrometry of order $\unit{10}{\micro\arcsec}$, this adaptive-optics assisted beam combiner (BC) observes two celestial objects simultaneously in dual-field mode. In this way, GRAVITY will revolutionize dynamical measurements of astronomical targets. In particular, its primary science goal is to probe physics in the Galactic Center close to the event horizon of the supermassive black hole, where certain effects predicted by the theory of general relativity are expected to take place. For this purpose, GRAVITY consists of three main components: the infrared wavefront sensors, the actual beam combiner instrument (BCI) and the laser metrology system. The latter determines the differential optical path difference (dOPD) between the science and the reference object resulting from their angular separation on sky. To that end, the metrology system measures internal dOPDs within the interferometer and subtracts them from the measured dOPDs between the targets to extract their dOPD on sky. Here, we present the general concept of the metrology which consists of a three-beam laser system injected via fiber optics to the BCI of GRAVITY and a homodyne detection scheme for three-beam interference using phase-shifting interferometry. In more detail, the metrology measures the phases of laser fringe patterns created in the pupil planes of the telescopes, from which the internal dOPDs can be derived.
The developed system fulfills the following physical requirements:
\begin{itemize}
\item \textbf{all beams of the interferometer are monitored in parallel},
\item \textbf{the total length of the instrument and the VLTI up to the primary space is covered by the system, which is the correct position for defining the astrometric baseline},
\item \textbf{the phases are measured with a nanometer-accuracy},
\item \textbf{the differential delay lines (DDLs) can be closed stably on basis of the metrology signal},
\item \textbf{the phases are tracked at a speed fast enough during the acquisition procedure, such that presetting times are not excessive}.
\end{itemize}
Subsequently, we describe the working principle of the metrology system followed by its implementation and a few concluding remarks.

\section{Working Principle}
\label{sec:work}  

\begin{figure} [H]
   \begin{center}
   \begin{tabular}{c} 
   \includegraphics[height=18cm]{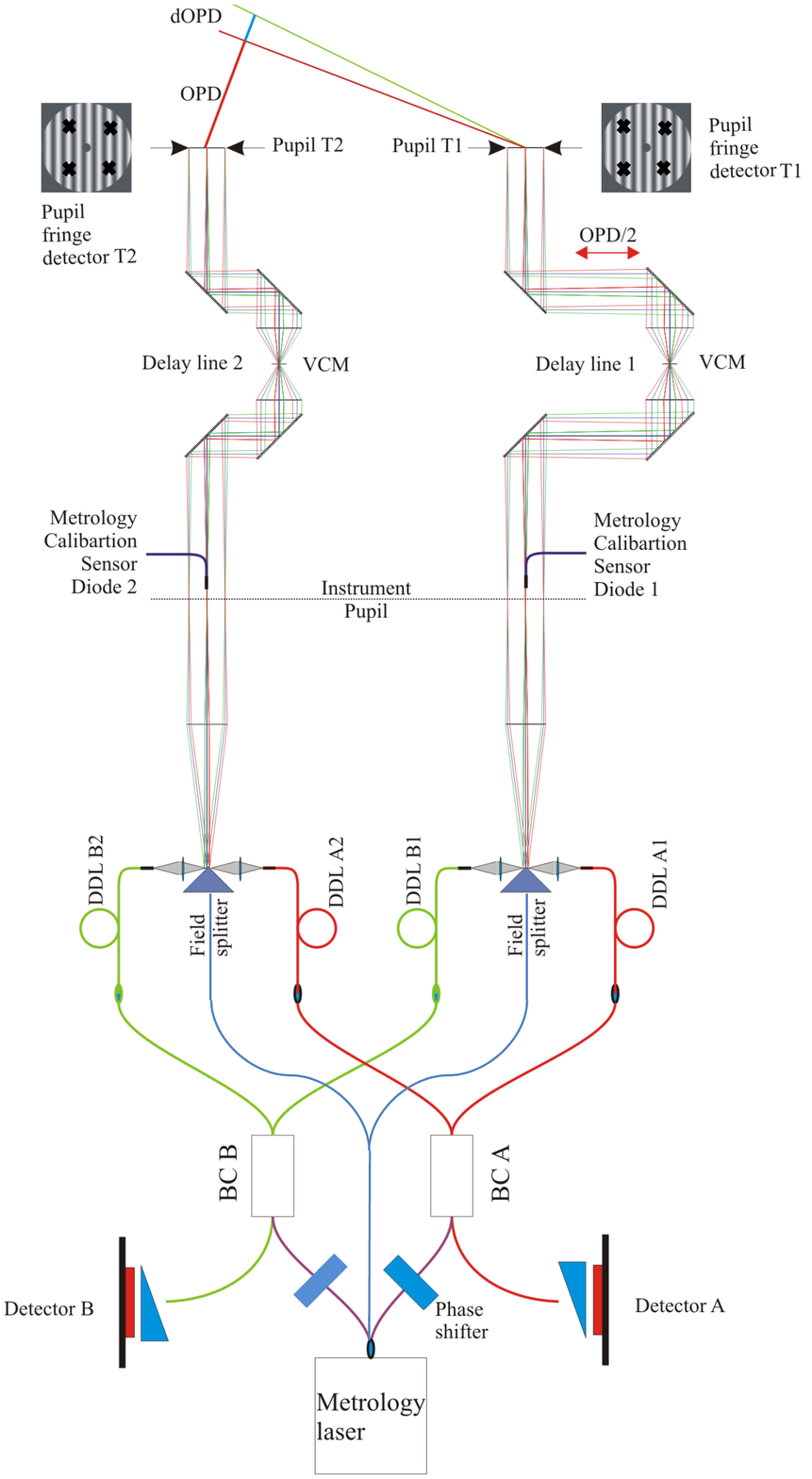}
   \end{tabular}
   \end{center}
   \caption[met] 
   { \label{fig:met} 
Working principle of the metrology system shown for two out of four telescopes. Laser light is split into equally faint beams and one high-power carrier beam. The faint beams are injected backwards into the BCs A and B of the two observed objects A and B. The carrier beam is split and superimposed to enhance the faint signal for detection. The beams trace the optical paths of the astronomical light back to the telescope pupils T1 and T2 where they interfere. For the homodyne detection each of the two faint beams is phase-shifted with an individual frequency. Lock-in amplifiers are coupled to each of the frequencies in order to disentangle the two fringe patterns between the faint beams and the carrier at each telescope. Subtracting the respective fringe phases from each other cancels out the carrier phase such that the actual fringe phase of interest between the faint beams is retrieved. The dOPD within the interferometer can be determined from the phase difference between the telescopes.}
\end{figure}

The working principle of the metrology system is demonstrated in Figure~\ref{fig:met}. The metrology traces all the optical paths from the beam combination of the astronomical light back to the telescopes. Since its measurements cover the full path of the instrument and the VLTI up to the primary mirror M1 and are performed in the pupil plane at fixed positions with respect to M1 after reflection on the latter, the definition of the astrometric narrow-angle baselines is essentially free of systematics~\cite{lacour}. In addition, the pupil is sampled at typical radii such that potential systematics due to differential focus is eliminated.
\\
The basic concept is to split laser light into three beams with fixed phase relations. Two of the beams are faint and injected backwards into the two BCs of the celestial objects. The third, high-power beam is overlaid on top of the two faint ones after they have passed fibers within the BCI. This approach is chosen in order to minimize inelastic scattering of the metrology light in the instrument fibers, which is backscattered onto the science detector. In more detail, fluorescence of the rare earth elements Holmium (Ho$^{3+}$) and Thulium (Tm$^{3+}$) as well as Raman emission in fluoride fibers, both excited by the metrology laser wavelength, can be observed in the detector band.
\\
Another advantage of this three-beam implementation is the suppression of non-common path effects from variations of high-power laser radiation. Their origin lies in the injection of the metrology light to the BCs which introduces optical paths that the science light does not travel, therefore the term ``non-common''. A certain part of these paths appears in the astrometric error budget and should not drift more than a few nanometers during an observation block. However, high laser powers can induce temperature fluctuations in the metrology injection fibers which result in variations of the fiber length implying such drifts~\cite{kok}. 
\\
At telescope level, the three light beams interfere in the pupil plane and form a fringe pattern which in principle carries the individual interferences between three beams. Temporal sampling of this fringe pattern with photodiodes allows for extracting the phase corresponding to the internal dOPD between the light paths of the reference and science object from one telescope. The phase of interest corresponds to the one of the interference between the faint beams which alone would be undetectably faint. Therefore, the third beam serves as an amplification for the fringe pattern. This is why, for the actual measurement of the phase between the faint beams, we analyze their individual interference with the third beam independently by phase-shifting each of them with a characteristic frequency. By means of lock-in amplifiers referenced to those frequencies the phases of the corresponding fringes can be extracted and in difference provide the desired internal dOPD between science and reference path while the phase of the high-power beam cancels out. Since we are only interested in the oscillating terms of the interference carrying the phase information, the AC signal, and not the average level of the oscillation, the metrology receivers are AC-coupled. This realization corresponds to a homodyne detection scheme in the sense that the lock-in amplifiers use the frequencies with which the metrology signals are modulated as reference.
Knowing the internal dOPDs for two telescopes encoded in the phases $\Phi_1$ and $\Phi_2$ allows for calculating the dOPD on sky between the observed objects as seen by this baseline from the difference of the astronomical fringe phases $\Psi_A$ and $\Psi_B$ measured on detector A and B via
\begin{equation}
dOPD = \frac{\lambda_A}{2\pi}\Psi_A - \frac{\lambda_B}{2\pi}\Psi_B + \frac{\lambda_L}{2\pi} \left( \Phi_2-\Phi_1 \right) - \frac{\lambda_L}{2\pi} \left( \delta_2 - \delta_1 \right) \hspace{0.5cm} .
\end{equation}
This formula includes the effective wavelengths of the light of star A and B, $\lambda_A$ and $\lambda_B$, as well as the wavelength of the metrology laser $\lambda_L$. In total, six baselines between four telescopes are addressed in that way. The contributions $\delta_1$ and $\delta_2$ denote the non-common path terms in the metrology injection which need to be calibrated for the data reduction~\cite{vincent} by swapping the light of two nearby stars between the BCs in order to determine the zero points of the metrology phases. These measurements are not performed online but in the preset of an observation. 

\section{Implementation}

\begin{figure} [H]
   \begin{center}
   \begin{tabular}{c} 
   \includegraphics[height=20cm]{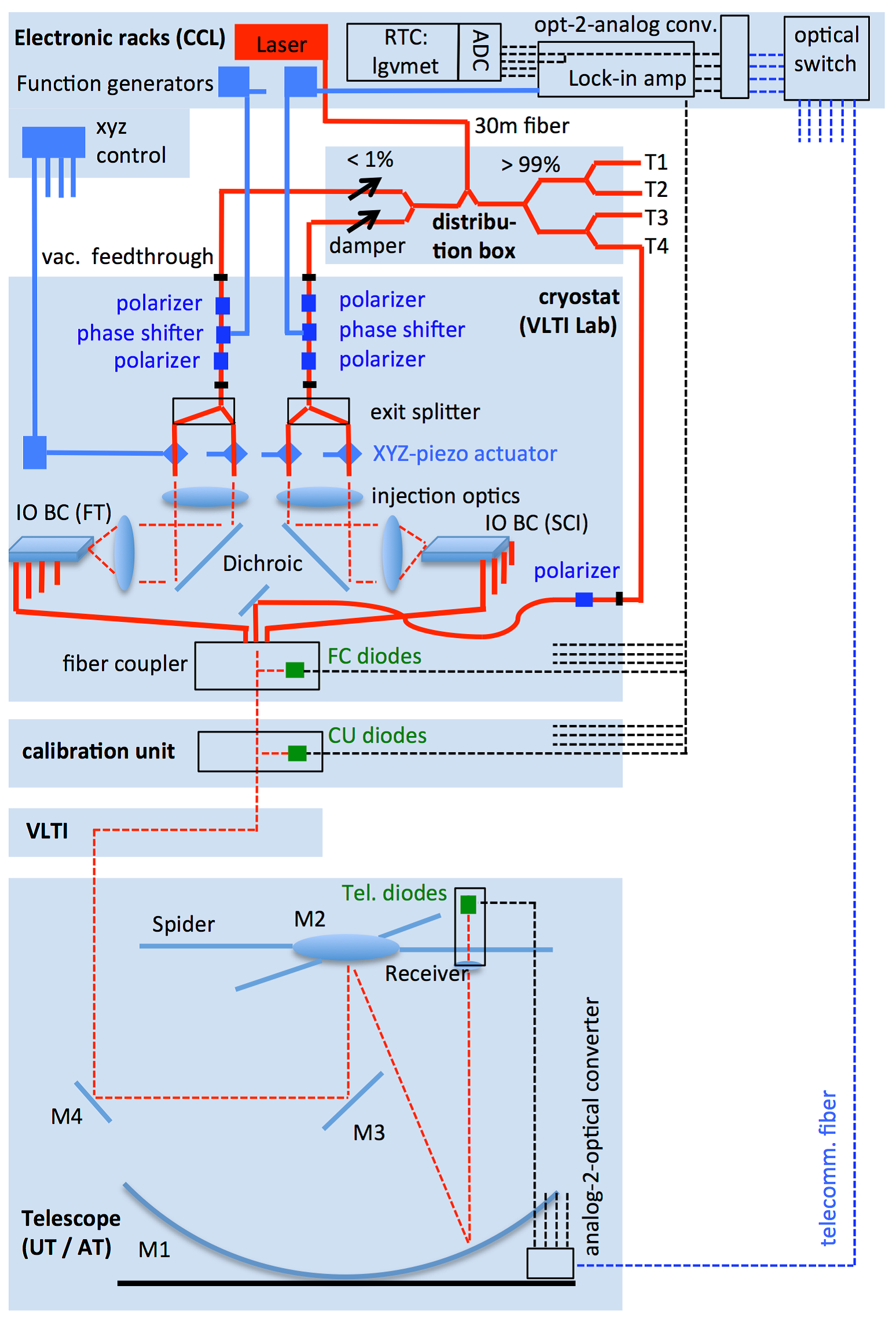}
   \end{tabular}
   \end{center}
   \caption[hardware] 
   { \label{fig:hardware} 
Sketch of the metrology hardware. For detailed description see Table~\ref{tab:components} and the text.}
   \end{figure}

The final system design consists of three main parts, namely the injection of the metrology laser light into the BCI, its detection in the latter and at telescope level as well as of the computing hardware and software. The GRAVITY software was presented in Reference~\citenum{burtscher} and~\citenum{ott} while this contribution focuses on the other aspects. The implemented hardware is sketched in Figure~\ref{fig:hardware} and the corresponding components are listed in Table~\ref{tab:components}. Here, we discuss selected key items.

\subsection{Laser}

The stabilized single frequency laser source from Menlo Systems produces light at a wavelength slightly below the astronomical K-band of $\unit{2.0}{\micro\meter}$ to $\unit{2.5}{\micro\meter}$ and has the following properties:
\begin{itemize}
\item power output: $>\unit{1}{\watt}$ at the end of a $\unit{30}{\meter}$ long fiber,
\item output wavelength: $\unit{1908}{\nano\meter}\pm\unit{1}{\nano\meter}$, stabilized to $\pm\unit{30}{\mega\hertz}$ absolute wavelength accuracy,
\item FWHM of the output wavelength distribution $<\unit{10}{\mega\hertz}$,
\item degree of linear polarization: $>100:1$ at fiber output,
\item polarization direction adjusted to slow axis of fiber and fiber connector key,
\item power output fluctuations (at the end of the $\unit{30}{\meter}$ fiber): $<0.5\%$ RMS  of the integrated laser power over a $\unit{30}{\milli\second}$ timescale,
\item total residual optical power, outside of the range $\unit{1907}{\nano\meter} \le \lambda \le \unit{1909}{\nano\meter}$: $<10^{-4}$ of total output power.
\end{itemize}

\subsection{Phase shifters}

The Photline phase shifters are electro-optic devices employing lithium niobate (LiNbO$_3$) crystals in which a change of refractive index can be induced by applying a voltage. Since this material is highly birefringent i. e. it has two transmission axes with different refractive indices, the linear polarized optical input should be aligned to one of these axes to maintain the linear polarization of the laser light otherwise elliptical polarization would be introduced when feeding both axes. However, due to many fiber connections in between the phase modulators and the laser source small misalignments can occur in the setup presented here. As a consequence, we spliced fiber polarizers from Thorlabs to the in- and output fibers of the phase shifters in order to correct for possible mismatches in front of and behind the devices. We apply two different phase-modulating frequencies for the science and reference metrology beam, both in the kilohertz-range.

\begin{table}[h]
\caption{List of the metrology hardware. The metrology system can be split into 3 subsystems: injection, detection and computing. Part of the injection components are located outside of the BCI, others are inside. } 
\label{tab:components}
\begin{center}       
\begin{tabular}{|l|l|l|} 
\hline
\rule[-1ex]{0pt}{3.5ex}  \textbf{Subsystem} & \textbf{Component} & \textbf{Function}  \\
\hline
\rule[-1ex]{0pt}{3.5ex}  Injection (outside): & Laser & Generates 1908nm laser light  \\
\cline{2-3}
\rule[-1ex]{0pt}{3.5ex}   & PM fibers & Guide laser light  \\
\cline{2-3}
\rule[-1ex]{0pt}{3.5ex}   & Distribution box & Contains fiber splitters and attenuators  \\
\cline{2-3}
\rule[-1ex]{0pt}{3.5ex}  & Vacuum feedthroughs & Feed metrology beams to BCI cryostat  \\
\hline 
\rule[-1ex]{0pt}{3.5ex}  Injection (inside): & Phase shifters & Phase-shift the beams for detection  \\
\cline{2-3}
\rule[-1ex]{0pt}{3.5ex}   & Polarizers & Define polarization at the injection points  \\
\cline{2-3}
\rule[-1ex]{0pt}{3.5ex}   & Exit splitters & Inject light to the two IO BCs  \\
\cline{2-3}
\rule[-1ex]{0pt}{3.5ex}   & XYZ-piezo actuators & Position the output fibers of the exit splitters  \\
\cline{2-3}
\rule[-1ex]{0pt}{3.5ex}   & Injection optics & Couple laser light to IO outputs  \\
\hline
\rule[-1ex]{0pt}{3.5ex}  Detection: & FC diodes & Detect metrology signals within the BCI \\
\cline{2-3}
\rule[-1ex]{0pt}{3.5ex}   & CU diodes & Mimic metrology signals from telescopes  \\
\cline{2-3}
\rule[-1ex]{0pt}{3.5ex}   & Telescope diodes & Detect metrology signals at telescopes  \\
\cline{2-3}
\rule[-1ex]{0pt}{3.5ex}   & Analog-2-optical converter & Converts analog signals to optical ones  \\
\cline{2-3}
\rule[-1ex]{0pt}{3.5ex}   & Optical switch & Connects to the selected telescopes  \\
\cline{2-3}
\rule[-1ex]{0pt}{3.5ex}   & Optical-2-analog converter & Converts optical signals to analog ones  \\
\cline{2-3}
\rule[-1ex]{0pt}{3.5ex}   & Lock-in amplifiers & Lock to phase-shifting frequencies  \\
\cline{2-3}
\rule[-1ex]{0pt}{3.5ex}   & Function generators & Drive phase-shifters, reference for lock-ins  \\
\hline
\rule[-1ex]{0pt}{3.5ex}  Computing: & ADCs & Read metrology signals for real-time computing  \\
\cline{2-3}
\rule[-1ex]{0pt}{3.5ex}   & LCU & Performs real-time computing  \\
\hline
\end{tabular}
\end{center}
\end{table}
 \begin{figure} [H]
   \begin{center}
   \begin{tabular}{c} 
   \includegraphics[height=5.5cm]{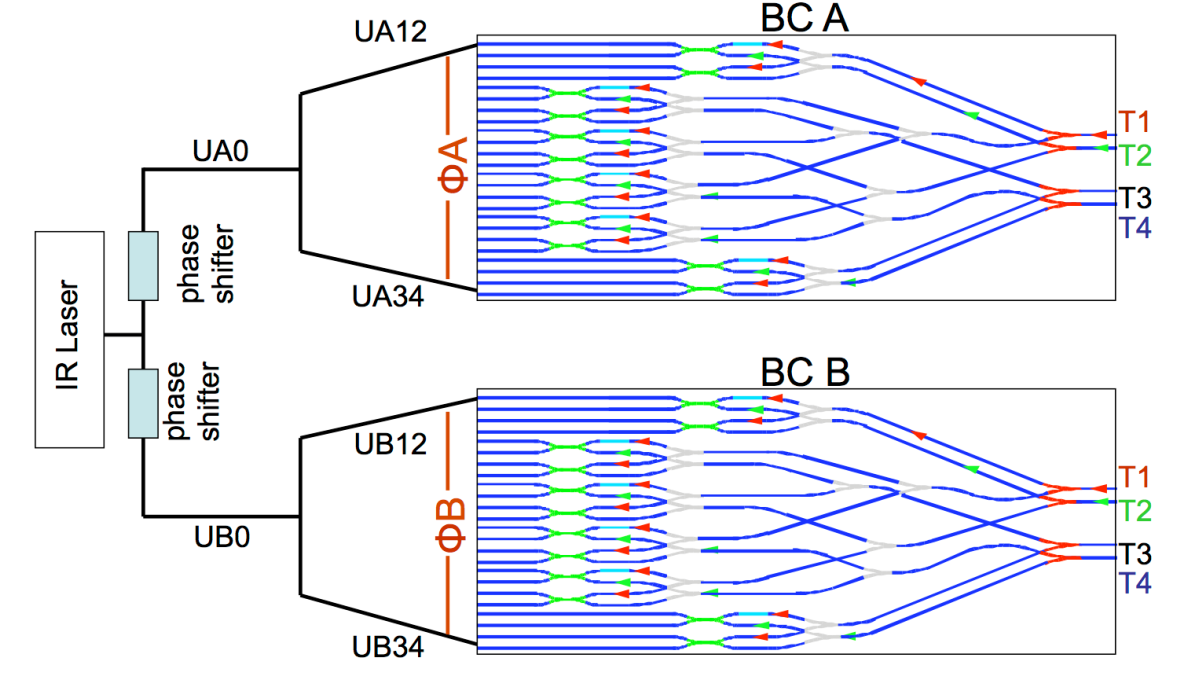}
   \end{tabular}
   \end{center}
   \caption[coupling] 
   { \label{fig:coupling} 
Coupling of the metrology light into the beam combiners BC A and BC B. The IR laser light is split into two beams, which can be phase-shifted independently. Before these beams enter the BCs another splitting is necessary in order to illuminate two of the 24 BC outputs such that all the four BC inputs T1, T2, T3 and T4 are fed with light. The non-common paths that are only passed by the metrology light not by the science beams are denoted $U$... . However, $U\!A0$ and $U\!B0$ have no influence on the astrometric calculations since they cancel in the phase subtraction between two telescopes which delivers the internal dOPD. Only the differences $(U\!A12-U\!A34)$ and $(U\!B12-U\!B34)$, corresponding to the phases $\phi_A$ and $\phi_B$, enter the budget.}
   \end{figure} 

\subsection{Fiber positioner units}

From this point the metrology light enters the two custom-made fiber positioner units which couple the laser beams to the BCs which are realized as integrated optics~\cite{io} (IO). In order to send light to all four telescopes, two of the 24 IO outputs (6 baselines x 4 phases = 24) need to be illuminated as demonstrated in Figure~\ref{fig:coupling}. For this reason the two beams are split again into two of equal intensity by means of LEONI fiber splitters which we denote as exit splitters. Their output fibers, not connectorized but cleaved, are glued into v-grooves on xyz-piezo actuators from SmarAct such that each fiber can be aligned remotely to the desired IO output via free-beam coupling optics. The positioning accuracy and the maximum position drift are smaller than $\unit{0.05}{\micro\meter}$.
\\
In order to minimize non-common path effects to a negligible level of the error budget the output fibers are kept as short as possible ($\approx\unit{5}{\centi\meter}$) and their maximum difference in length is $\unit{1}{\milli\meter}$. Furthermore, those fibers are single-mode (SM) but not polarization-maintaining (PM) in contrast to the other parts of the metrology chain including the input fibers of the exit splitter. This design leads to smaller changes of the output polarization state from bending of the fibers by actuators than for PM fibers as shown in laboratory tests. The short SM fibers maintain the polarization state nevertheless because their beat length is much larger than the actual fiber length other than for the birefringent PM fibers.

\subsection{Injection Optics}

\begin{figure} [H]
   \begin{center}
   \begin{tabular}{c} 
   \includegraphics[height=7cm]{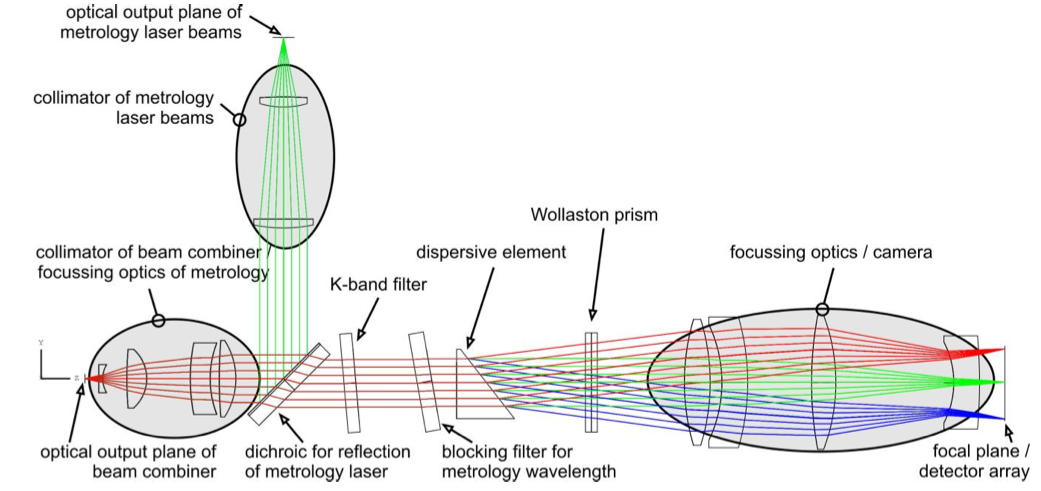}
   \end{tabular}
   \end{center}
   \caption[spectro] 
   { \label{fig:spectro} 
Coupling optics for the metrology injection. The output beams from the exit splitters are guided by a collimator to a $\unit{45}{\degree}$-dichroic beam splitter where they join the science light. Via another collimator the metrology light is then coupled to the waveguides of the IO. In order to minimize back-reflections the output surface of the BC chips is coated. In this respect, any back-reflected light of the metrology wavelength $\unit{1908}{\nano\meter}$ travelling through the dichroic in the direction of the detector is blocked by two filters with a combined OD $>16$. The fraction of the injected metrology light which passes through the dichroic is captured in a beam dump not drawn here. }
   \end{figure}

The output of the exit splitters is then guided through injection optics illustrated in Figure~\ref{fig:spectro}. Essentially, they consist of collimators, a dichroic, a beam dump as well as a filter and are part of the GRAVITY spectrometers~\cite{straubi}. A demagnification factor of 1.897 is realized to match the Gaussian modes of the metrology fibers with mode field radius (MFD) of $\unit{12.9}{\micro\meter}$ and the IO with MFD $\unit{6.8}{\micro\meter}$ for the metrology wavelength. As such, the output fibers of the exit splitters need to be positioned at a distance of $\unit{340}{\micro\meter}$ to each other in order to map the distance between the IO channels of $\unit{180}{\micro\meter}$. The dichroic reflects the metrology light onto the BC while the passing fraction of the light is captured by the beam dump. Two filters behind the dichroic block the back-reflected laser light of $\unit{1908}{\nano\meter}$ going in the direction of the detector.

\subsection{Detection}

After the injection of the two metrology beams to the BCs they travel all the paths of the astronomical light backwards up to the telescopes where their interference pattern is sampled in the pupil plane by four receivers per telescopes. More precisely, the corresponding photodiode boxes are mounted at the four spider arms above the primary mirror M1 at the entrance pupil of the secondary mirror M2. This design is realized at the Unit Telescopes (UTs) of the VLTI as well as at the Auxiliary Telescopes (ATs). As all the receivers are installed at the same radius the average phase is independent of tip and tilt. Before, this radius was optimized for the average to be focus-independent. 
\\
The receivers are equipped with optics that focus the incoming light onto the sensitive area of the photodiode of $\unit{0.25}{\milli\meter}$ in diameter by a Thorlabs lens and suppress background light by a line filter from Filtrop. Via electronics built here at MPE, the input signals detected by the diodes from Laser Components are then amplified in two stages. The AC coupling mentioned in section~\ref{sec:work} is implemented after the first amplification stage. In addition to those telescope detectors, four receivers are installed within the BCI as internal reference, one per telescope. These diodes are located in the fiber couplers~\cite{fc} (FC) and also pick up the metrology light in the pupil, but with a multimode fiber glued onto the back of a lens. The picked-up light is then detected by means of a fiber-to-photodiode coupler from OZ Optics. A similar set of four diodes is placed in the calibration unit~\cite{cu} (CU) of the instrument which mimics the telescope signals when operating GRAVITY in stand-alone or calibration mode. 
\\
The final step of the metrology detection is the processing by the Femto lock-in amplifiers, two per detected analog voltage signal, in order to lock onto the two different phase-shifting frequencies of the science and reference metrology beam as outlined above. Function generators from Keysight induce the phase modulation by linear voltage ramps which at the same time provide the reference frequency for the lock-in detection. The amplifier itself produces two signals proportional to the sine and cosine of the phase between the carrier beam and one of the faint beams. In total we therefore record 80 signals coming from 4 telescopes with 4 receivers each plus 4 internal ones, so a total of 20 detectors. Each of the receivers is locked by 2 amplifiers to extract the two phases of each of the faint beams with the carrier beam, which in turn produce 2 signals such that in the end 20 detectors produce 80 signals. These 80 voltages are read via analog-to-digital converter (ADC) cards into the real-time computer (RTC). 
\\
For the latter we use a dedicated local control unit (LCU) called lgvmet. The phase information required for the determination of the internal dOPDs can be calculated from those signals. This metrology design allows for nanometer-precision measurements in GRAVITY's astrometric mode. Concluding remarks are presented in the following last section.

\section{Conclusions}

GRAVITY aims for precision-astrometry of order $\unit{10}{\micro\arcsec}$ by observing two celestial objects in dual-field mode. For doing so, the dOPD corresponding to the angular separation between the two needs to be determined by tracking the internal dOPDs within the interferometer.
These internal measurements are performed by a dedicated metrology system consisting of a three-beam laser system. A homodyne detection scheme is used to analyze the resulting three-beam interference by means of lock-in amplifiers and phase-shifting. The main physical requirements for the operation of the metrology are the fringe detection in primary mirror space, a minimized light level of the metrology within the BCI and errors on the phase measurements of nanometer-level. The fringe detection in the pupil plane of the telescopes is given by the metrology design, while the low light level and the required phase error need to be balanced against each other. The low light level is needed in order to minimize non-common path effects and backscattering of the metrology light onto the science detectors. However, the phase error is determined by the inverse SNR which in turn decreases with lower light levels and thus increases the error on the phase measurement. In autumn 2015, GRAVITY was integrated in the VLTI structure and since then has been undergoing various perfomance tests in combination with technical work. In this respect, the detection elements required for the metrology system were installed at all eight telescopes. In particular, the receivers were aligned optically to maximize the detected metrology signal and radially positioned by means of a laser tracker with an accuracy of less than $\unit{1}{\milli\meter}$ for a focus-independent phase average over the illuminated pupil. At the time of writing the project is still in its commissioning phase and we are optimizing the metrology performance on this basis, in particular investigating on the lowest light levels and SNRs respectively which the metrology system can operate on.

 \begin{figure} [H]
   \begin{center}
   \begin{tabular}{c} 
   \includegraphics[height=6.5cm]{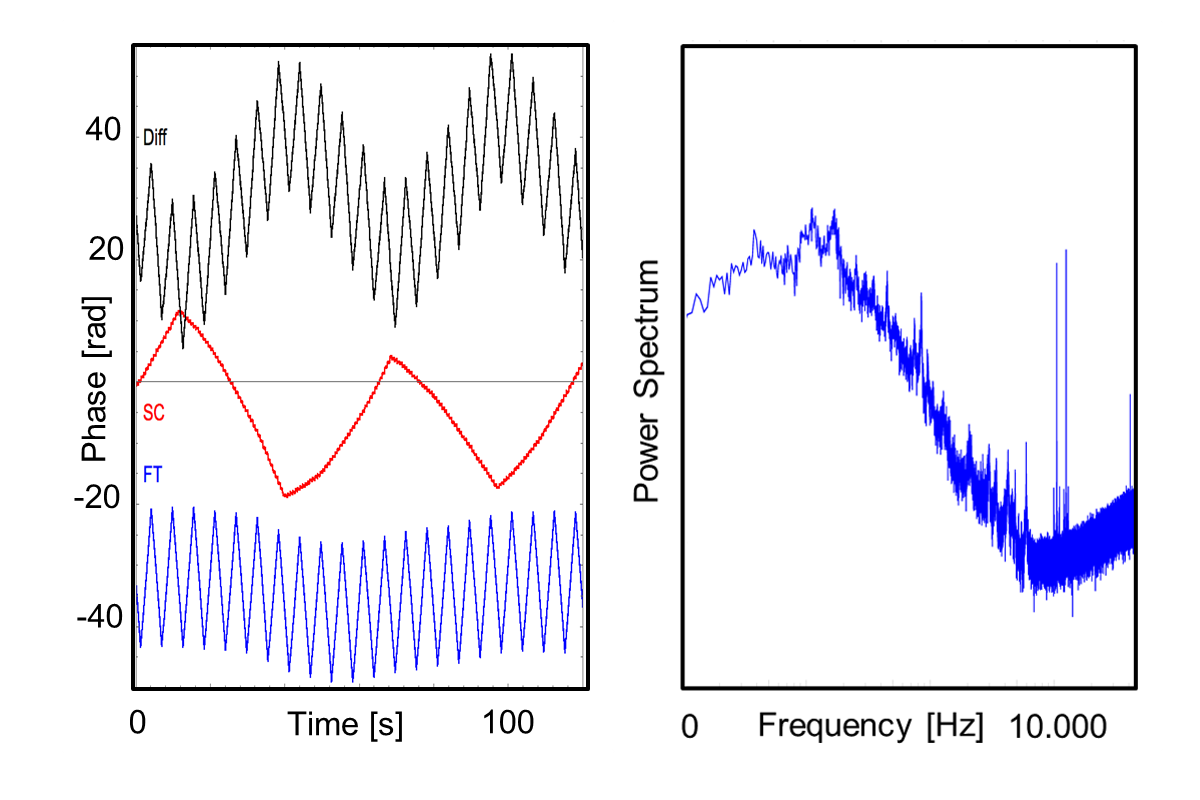}
   \end{tabular}
   \end{center}
   \caption[psd] 
   { \label{fig:psd} Demonstration of the detection scheme. On the left side the phase modulation of the two faint beams is visualized in red and blue as well as the phase difference between them in black. The beams are modulated with two distinct frequencies in order to detect their individual interference with the carrier beam separately by means of lock-in amplifiers referenced to the modulations. The phase difference is the quantity of interest since it encodes the internal dOPD. The power spectrum of the metrology signal seen by the receivers is shown on the right side with logarithmic frequency scale. The low frequencies are dominated by the laser power noise from the carrier beam. This noise term is high-pass filtered by the receiver such that the power spectrum consists of the receiver noise and the two modulation peaks at the high frequencies. The modulation frequencies are at $\unit{9}{\kilo\hertz}$ and $\unit{11}{\kilo\hertz}$. For better visualization they are chosen differently in the left plot.
}
   \end{figure} 

\bibliography{report} 

\bibliographystyle{spiebib} 

\end{document}